\documentclass[titlepage,12pt]{article}
\usepackage{graphicx}
\usepackage[myheadings]{fullpage}
\usepackage{pmetrika}
\usepackage{submit}
\usepackage{bm,amssymb,amsmath,amsthm,dsfont}
\usepackage{verbatim}
\usepackage{multirow}
\usepackage{appendix}
\usepackage{float} 
\usepackage[margin = 1in]{geometry}
\usepackage{setspace}
\usepackage{paralist}
\usepackage[flushleft]{threeparttable}
\usepackage{textcomp}
\usepackage{tikz,makecell}

\usetikzlibrary{positioning}
\usepackage[natbibapa]{apacite}
\usepackage{hyperref}
\usepackage{color}
\usepackage{latexsym,color}
\usepackage{url}
\usepackage{caption,subcaption}
\usepackage{wrapfig}
\usepackage{framed}
\usepackage{amsfonts}
\usepackage{rotating}
\usepackage{cases} 
\usepackage{xr}
\usepackage{booktabs}

\DeclareMathOperator*{\argmax}{argmax}

\DeclareMathOperator{\cov}{Cov}

\newtheorem{lemma}{Lemma}
\newtheorem{theorem}{Theorem}

\newtheorem{procedure}{Procedure}

\begin{document}

\begin{titlepage}

\title{Accurate Assessment via Process Data}

\author{Susu Zhang, Zhi Wang, Jitong Qi, Jingchen Liu and Zhiliang Ying}
\affil{Columbia University}

\vspace{\fill}\centerline{\today}\vspace{\fill}

\linespacing{1}


\end{titlepage}

\setcounter{page}{2}
\vspace*{2\baselineskip}

\RepeatTitle{Accurate Assessment via Process Data}\vskip3pt

\linespacing{1.75}
\abstracthead
\begin{abstract}
Accurate assessment of student's ability is the key task of a test. Assessments based on final responses are the standard. As the infrastructure advances, substantially more information is observed. One of such instances is the process data that is collected by computer-based interactive items and contain a student's detailed interactive processes. In this paper, we show both theoretically and with simulated and empirical data that appropriately including such information in assessment will substantially improve relevant assessment precision. 

\begin{keywords}
Process data; ability estimation; automated scoring; Rao-Blackwellization\end{keywords}
\end{abstract}\vspace{\fill}\pagebreak

\section{Introduction}

The main task of educational assessment is to provide reliable and valid estimates of students' abilities based on their responses to test items. Much of the efforts in the past decades focused on the item response theory (IRT) models, the responses of which are often dichotomous (correct/incorrect), polytomous (partial score), and generically discrete (e.g.~multiple choice item).
The rapid advancement of information technology has enabled the collection of various sorts of process data from assessments, ranging from reaction times on multiple choice questions to the log of problem-solving behavior on computer-based constructed-response items. In particular, the sequence of actions performed by test takers to solve a task, which document the processes that the test takers go through to solve a problem, can contain valuable information on top of final responses, that is, dichotomous or polytomous scores on how well the task was completed. The analysis of process data has recently gained strong interest, with a wide range of model- and data-driven methods proposed to understand the types of strategies that contribute to successful/unsuccessful problem-solving, identify the behavioral differences between observed and latent subgroups, and assess the proficiency on the trait of interest, et cetera \citep[e.g.,][]{he2016analyzing,lamar2018markov, liu2018analysis,xu2018latent}.

The emergence of process data provide the psychometric community with a great opportunity to develop cutting edge research and at the same time bring forward a great challenge. Process data indeed contain rich information about the students. Much of the literature focuses on developing new research directions.
In this paper, we take a different angle and try to answer the question how existing research could benefit from the analysis. 
In particular, we develop a method to incorporate information in process data to the scoring formula. There are two key features to consider: reliability and validity.  
For reliability, we show that the process-data-based assessment is significantly more accurate than that based on the IRT models. In particular, we demonstrate through a real data analysis that the process-data-based scoring rule yields much higher reliability that that of the IRT-model-based ability estimates. Score based on a single process data item could be as accurate as that of three IRT-based scores.
Furthermore, we also provide a theoretical framework under which process-data-based scores are guaranteed to yield more accurate estimates of students' abilities, of course, under certain conditions.
Reliability, on the other hand, is a more complex problem. Process data records the entire problem-solving process that reveals different aspects of a student. It is unclear which part of the process is related to the particular ability of interest. It is conventionally up to the domain experts to identify construct-relevant process features and derive scoring rules. 
Such an approach is costly and cannot be scaled up. 
Our approach considers an automated scoring system of process data. Features are extracted through an exploratory analysis that typically lacks interpretation. We take advantage of the IRT scores that helps us to guide scoring rules to yield a valid test score. The entire procedure does not require particular knowledge of the item design.

Process data is often in a format not easy to directly incorporate to analysis. We preprocess the data by embedding each response process to a finite dimensional vector space. There are multiple methods to fill this task, including $n-$gram language modelling \citep{he2016analyzing}, sequence-to-sequence autoencoders \citep{tang2019seq2seq}, and multidimensional scaling \citep{tang2019mds}. In this paper, we use feature extracted from multidimensional scaling. 
	
In the psychometrics literature, plenty of research has been conducted on the use of additional information, such as response times, to improve measurement accuracy \citep[e.g.,][]{bolsinova2018improving}. To the authors' best knowledge, this is the first piece of work to use problem-solving log data to improve measurement accuracy. A literature that is remotely related to the currently work is the automated scoring systems for constructed responses. Extensive research has been on on automated scoring of essays, which aims at producing essay scores comparable to human scores based on examinees' written text \cite[e.g., ][]{page1966imminence,attali2006automated,foltz1999automated,rudner2006evaluation}. Other than essay scoring, automated scoring engines have been developed for scenario-based questions in medical licensure exams \citep{clauser2000generalizability}, constructed-response mathematics problems \citep{fife2013automated}, speaking proficiency exams \citep{evanini2015automated}, and prescreening for post-traumatic stress disorder based on participant self-narratives \citep{he2017automated,he2019combining}. 
Many of these systems were shown to produce comparable scores to expert ratings. Readers are referred to \cite{bejar2016automated} and \cite{rupp2018designing} for comprehensive reviews of the history, applications, conceptual foundations, and validity considerations of automated scoring systems. 

The proposed approach differs from most automated scoring systems in its objective. Whereas automated scoring systems are often designed to reproduce expert- or rubric-derived scores in an automated and standardized manner, the purpose of the proposed Rao-Blackwellization approach is not to reproduce the final scores but to refine the latent trait estimates based on original final scores with the additional information from the problem-solving processes.

The rest of the paper is organized as follows. Section 2 describes the statistical formulation, and the proposed method for process-based score refinement is introduced. Theoretical results on mean squared error (MSE) reduction in latent trait estimation are presented. Section 3 reports results on simulation studies that verify the theoretical findings. Section 4 describes an empirical example on the problem-solving in technology-rich environments (PSTRE) assessment in the 2012 Programme for the International Assessment of Adult Competencies (PIAAC) survey, where the proposed method is compared to original response-based scoring in several aspects. A discussion of the implications and limitations is provided in Section 5.

\section{Latent Trait Estimation with Processes and Responses}

We start with presenting a generic framework for the proposed approach, and a more specific illustrative example is provided afterwards. Section \ref{sec:formulation} describes a statistical formulation that the proposed approach is built upon. Section \ref{sec:procedure} describes the proposed Rao-Blackwellization approach for process-based latent trait measurement, as well as the theoretical results on MSE reduction. Section \ref{sec:illustration} presents a more detailed illustration, which exemplifies one way the generic approach can be implemented in practice.

\subsection{Statistical Formulation}\label{sec:formulation}

Consider a test of $J$ items that is designed to measure a latent trait, $\theta$.	
For an examinee, on each item $j$, both the final item response and the action sequence for problem-solving are recorded. Denote the item response by $Y_{j}$, which can be a polytomous score ranging between $0$ and $C_j$ representing different degrees of task completion. Further denote the action sequence by $\boldsymbol S_{j} = (S_{j1}, \ldots, S_{jL_{j}})$, where $L_{j}$ is the total number of actions performed on the item, and $S_{jl}$ is the $l$th action. 

We consider the case where the action sequences record problem-solving details and thus contain at least as much information as the final outcomes. In this case, the final item response can be derived from the action sequence through a deterministic scoring rule $f$ such that $Y_{j} = f(\mathbf S_{j}).$ 
Further suppose that the final responses to the $J$ items are conditionally independent given $\theta$ and follow some item response function \citep[e.g.,][]{lord2012applications}, 
\begin{equation*}
    P(Y_{j} = y_j \mid \theta, \boldsymbol\zeta_j),
\end{equation*}
where $\boldsymbol\zeta_j$ is the parameters of item $j$.

For the present purpose of latent trait estimation, we assume that the item parameters ($\boldsymbol\zeta_j$s) have been calibrated and only the latent trait $\theta$ is unknown. Denote the pre-calibrated parameters of item $j$ by $\hat{\boldsymbol\zeta}_j$.
The latent trait $\theta$ for each individual can be estimated based on the response from one or more items.
 Commonly used latent trait estimators include the maximum likelihood estimator (MLE), where 
\begin{equation}\label{MLE}
    \hat\theta^{MLE} = \argmax_{\theta} \sum_{j} \log(P(Y_{j} = y_{j}\mid \theta, \hat{\boldsymbol\zeta}_j)),
\end{equation}
the Bayesian expected a posteriori (EAP) and Bayesian modal estimators (BME), i.e.,
\begin{equation}\label{EAP}
    \hat\theta^{EAP} = E[\theta\mid \mathbf Y] \text{  ~ and ~  } 
    \hat\theta^{BME} = \argmax_\theta P(\theta\mid \mathbf Y),
\end{equation}
where $P(\theta\mid \mathbf Y) \propto p(\theta) \prod_{j} P(Y_{j} = y_{j}\mid \theta,\hat{\boldsymbol\zeta}_j)$ with $p(\theta)$ being the prior distribution \citep[e.g., ][]{kim1993ability}.

We aim at refining the $\theta$ estimators with a procedure that makes use of process data. Since action sequences are in non-standard format, instead of working directly with $\boldsymbol S_{j}$, we work with the $K-$dimensional numerical features extracted from $\boldsymbol S_{j}$, denoted $\mathbf X_{j} = (X_{j1}, \ldots, X_{jk}) \in \mathbb{R}^K.$ There are no restrictions on the feature extraction method except that the produced features $\mathbf{X}_{j}$ must preserve the full information on the final response $Y_{j}$, in other words, $\sigma(Y_{j}) \subseteq \sigma(\mathbf X_{j})$, where $\sigma(\cdot)$ denotes the $\sigma-$algebra generated by the random variable. Intuitively, this requires the extracted features to preserve full information about the final score so that they can perfectly predict them. 
Since final outcomes are deterministically derived from response processes, they can always be added into extracted features to guarantee $\sigma(Y_j) \subseteq \sigma(\mathbf X_j)$. 
Feature extraction methods such as $n$-gram language modelling \citep[e.g.,][]{he2016analyzing,qiao2018data}, multidimensional scaling \citep[MDS; ][]{tang2019mds}, and recurrent neural network-based sequence-to-sequence autoencoders \citep{tang2019seq2seq}, which have documented performance in terms of near-perfect final response prediction, can be applied in practice.

\subsection{Procedure}	\label{sec:procedure}

Let $\mathbf X = (\mathbf X_1, \ldots, \mathbf X_J)$ denote the process features from all $J$ items and $\mathbf X_{-j}$ the process features from the $J-1$ items excluding item $j$. 
Denote the latent trait estimate based on all $J$ final responses by $\hat \theta_{\mathbf Y}$, which can be obtained through the estimators in Equations \eqref{MLE} or \eqref{EAP}. Further, let $\hat \theta_{Y_j} $ be the estimator derived from a single response outcome $Y_j$, for instance, using the EAP estimator in \eqref{EAP}. 

The final response-based trait estimator, $\hat \theta_{\mathbf Y}$, can be refined by the following procedures that incorporate process features. The new estimator is denoted by $\hat \theta_{\mathbf X}$.

\begin{procedure}[Construction of process-based estimator]\label{ch4:procedure:loo}~
	\begin{enumerate}
		\item For each $j = 1,\ldots, J:$
		\begin{itemize}
		    \item Regress $\hat \theta_{Y_j} $ on $\mathbf X_{-j}$ to obtain $T_{\mathbf X_{-j}} = E[\hat \theta_{Y_j} | \mathbf X_{-j}]$.
		    \item Regress $\hat \theta_{\mathbf Y}$ on $T_{\mathbf X_{-j}} $ and $Y_j$ to obtain $\hat \theta_{\mathbf X_{-j}} = E[\hat \theta_{\mathbf Y} | T_{\mathbf X_{-j}} , Y_j]$.
		\end{itemize}
		\item Compute the overall process-based estimator, $\hat \theta_{\mathbf X} = \frac{1}{J} \sum_{j=1}^{J} \hat \theta_{\mathbf X_{-j}} $.
	\end{enumerate}
\end{procedure}

In practice, the explicit distributions of $\hat\theta_{Y_j} | \mathbf X_{-j}$ and $\hat\theta_{\mathbf Y} | T_{\mathbf X_{-j}}, Y_j$ are unknown. The two conditional expectations, $E[\hat\theta_{Y_j} | \mathbf X_{-j}]$ and $E[\hat\theta_{\mathbf Y} | T_{\mathbf X_{-j}}, Y_j]$ in Procedure \ref{ch4:procedure:loo} can be approximated on finite samples using generalized linear models. Alternatively, deep neural networks can be fitted to capture the nonlinear relationships. Although, $J$ regressions are required for both steps, the implementation can be easily paralleled to make it computationally efficient.

The rest of this subsection illustrates the proposed procedure under a specific setting. Consider a test of $J$ binary items administered to $N$ respondents. For respondent $i$ and  item $j$,  let $\boldsymbol s_{ij}$ and $Y_{ij} \in \{0, 1\}$ denote the response process and the response outcome, respectively. 
Suppose that the response outcomes follow a two parameter logistic model \citep[2PL;][]{birnbaum1968models},
\begin{equation}
\text{logit} \left(P\left(Y_{ij} = 1 | \theta_i \right)\right) = a_j (\theta_i - b_j).
\end{equation}
The following steps provide a roadmap to implement Procedure \ref{ch4:procedure:loo}.
\begin{enumerate}[(1)]
	\item IRT parameter estimation: Fit the 2PL model on the binary outcomes $ \{Y_{ij}: i =1, \ldots, N, j=1,\ldots, J \}$ to obtain the item parameter estimates $\{\hat {\boldsymbol \zeta}_j = (\hat a_j, \hat b_j): j = 1, \ldots, J \}$ using marginal maximum likelihood estimation.
	
	\item Process feature extraction:  For each item $j$, extract features $\mathbf X_{1j},  \ldots, \mathbf X_{Nj}$ from the problem-solving processes $\boldsymbol S_{1j}, \ldots, \boldsymbol S_{Nj}$. The MDS method \citep{tang2019mds} or the action sequence autoencoder \citep{tang2019seq2seq} can be used for this step.
	
	\item Response-based (baseline) latent trait estimation:  We can choose from the commonly used estimators described in \eqref{MLE} or \eqref{EAP}. For each respondent $i$ and item $j$, get the single-item estimate $\hat \theta_{i, Y_j}$ based on $Y_{ij}$ and $\hat {\boldsymbol \zeta}_j$. Additionally, based on examinee $i$'s final responses to all $J$ items, $\mathbf Y_i = (Y_{i1}, \ldots, Y_{iJ})$, and $\hat{\boldsymbol \zeta} = (\hat{\boldsymbol \zeta}_1, \ldots, \hat{\boldsymbol \zeta}_J)$, estimate $\hat \theta_{i, \mathbf Y}$.
	
	\item First conditional expectations: For each $j$, fit a regression $\hat\theta_{Y_j} \sim \mathbf X_{-j}$ on $\{(\mathbf X_{i(-j)}, \hat \theta_{i, Y_j}): i = 1, \ldots, N\}$ to approximate $E[\hat\theta_{Y_j} | \mathbf X_{-j}]$ and calculate the fitted values $\{T_{i, \mathbf X_{-j}}: i = 1, \ldots, N\}$. For example, one can use ridge regression \citep{hoerl1970ridge} with shrinkage parameter selected by cross validation.  

	\item Second conditional expectations: For each $j$, fit another regression $\hat \theta_{\mathbf Y} \sim (T_{\mathbf X_{-j}}, Y_j)$. One simple choice is ordinary least squares with $(1, T_{\mathbf X_{-j}}, Y_j, T_{\mathbf X_{-j}}Y_j)$ as predictors, where $T_{\mathbf X_{-j}}Y_j$ is the interaction term.
	
	\item Averaging step: The average of the fitted values $\{\hat \theta_{i, \mathbf X_{-j}}: j = 1, \ldots, J\}$ in step (5) is the final process-based trait estimate, $\hat \theta_{i, \mathbf X}$, for respondent $i$.
\end{enumerate}

\subsection{Theoretical Analysis}\label{ch4:sec:theory}
The proposed procedure can improve latent trait estimation under the assumptions presented below. The first assumption requires the conditional expectation of $\hat \theta_{Y_j}$ given $\theta$ to be monotonically increasing. This assumption is satisfied by well-designed cognitive items and latent trait estimators. 
\begin{enumerate}
	\item[A1.] (Monotonicity assumption) $m_j(\theta) = E\left[\hat\theta_{Y_j} | \theta\right]$ is monotone in $\theta$ and has a finite second moment.
\end{enumerate}
Secondly, we assume that the response outcome of item $j$ is correlated with individual's behaviors on other items only through the measured trait $\theta$ and not through other latent or observed traits. Since the process features can include rich information other than the measured trait, this local independence assumption requires $Y_{j}$ to be ``good'',  in the sense that no construct-irrelevant, persistent traits affect final performance. In other words, measurement error comes of sources of random, instead of systematic error \citep{american2014standards}. For example, the process features $\mathbf X_{-j}$ may reflect a respondents' computer usage habits, such as whether they tend to use double- or single-clicking on buttons. However, the final score, $Y_j$, shall not differentiate individuals with different clicking habits, as long as they have the same level of $\theta$. We do allow $Y_j$ to be very ``rough'' measurements, in other words, the measurement error can be large, as long as it is due to random instead of systematic variations. Similarly, $\hat \theta_{Y_j}$ can be biased and can have large standard error, as long as the monotonicity assumption (A1) is satisfied.
\begin{enumerate}
	\item[A2.] (Local independence assumption) Given latent trait $\theta$,  $ Y_j$ and  $\mathbf X_{-j}$ are independent.
\end{enumerate}

Finally, we consider the distribution of the process features, $\mathbf X_{-j}$, given the measured trait. Note that the problem-solving processes can depend on traits other than $\theta$. These unobserved, construct-irrelevant traits are assumed random and integrated out from the probability density, thus resulting in the conditional density of $\mathbf X_{-j}$ given $\theta$. We impose the usual exponential family assumption on process features for technical development. This is equivalent to assuming the existence of a unidimensional sufficient statistic of $\theta$ of sample size \citep{lehmann2005testing}. The natural parameter $\eta_j(\theta)$ is assumed to be monotone so that there is no identifiability issue for $\theta$. 

\begin{enumerate}
	\item[A3.] (Exponential family assumption) The probability density function for features $\mathbf X_{-j}$ takes the following form
	\begin{equation}
	f(\mathbf X_{-j} | \theta) = \exp \left\{\eta_j\left(\theta\right) T_j(\mathbf X_{-j}) - A_j(\theta)\right\} h_j(\mathbf X_{-j}),
	\end{equation}
	where $T_j(\mathbf X_{-j})$ is a sufficient statistic for $\theta$ and the natural parameter $\eta_j\left(\theta\right)$ is monotone in $\theta$ with a finite second moment.
\end{enumerate}

Theorem \ref{ch4:thm:sufficient_stat} shows that the first step of our proposed procedure can summarize extracted features into sufficient statistics.
\begin{theorem}\label{ch4:thm:sufficient_stat}
	Under Assumptions A1--A3, $T_{\mathbf X_{-j}}$ is a sufficient statistic of $\mathbf X_{-j}$ for $\theta$.
\end{theorem}

Based on the sufficiency of $T_{\mathbf X_{-j}}$, we can further show that $\hat \theta_{\mathbf X}$ reduces the MSE of $\hat \theta_{\mathbf Y}$, as stated in Theorem \ref{ch4:thm:var_reduction}. The proof of this result uses the Rao-Blackwell theorem \citep{blackwell1947conditional, casella2002statistical} and also shows that every  $\hat \theta_{\mathbf X_{-j}}$ produced by step 2 of the procedure removes conditional variance and improves $\hat \theta_{\mathbf Y}$ in terms of MSE. The proofs of Theorem \ref{ch4:thm:sufficient_stat} and Theorem \ref{ch4:thm:var_reduction} are provided in the Appendix.

\begin{theorem}\label{ch4:thm:var_reduction}
	If assumptions A1-A3 hold for all $J$ items,  then
	\begin{equation}
	E[(\hat \theta_{\mathbf X} - \theta)^2|\theta]\leq E[(\hat \theta_{\mathbf Y} - \theta)^2|\theta] \quad {\it for \; every \; \theta}.
	\end{equation}
\end{theorem}

Putting Theorem 2 in the psychometrics context, the MSE reduction of $\theta$ estimator translates to the reduction of standard error of measurement. 
The proposed approach can be applied in practice to derive more reliable scores (latent trait estimates) for individuals in assessments. Alternatively, by using the procedure to incorporate process data, one is able to achieve comparable measurement precision to traditional outcome-based scoring with fewer items. 

\section{Simulations}\label{sec:sim}
In this section, simulation studies were conducted to compare outcome- and process-based estimators of the latent trait. 
\subsection{Experiment Settings}
We generated respondents' latent trait $\theta_1, \ldots, \theta_N$ independently from the standard normal distribution. Given respondent $i$ and item $j$, the response outcome $Y_{ij}$ followed a Rasch model \citep{rasch1960models},
\begin{equation}\label{eq:rasch}
\text{logit} \left(P\left(Y_{ij} = 1 | \theta_i \right)\right) = \theta_i - b_j.
\end{equation}
To generate the response processes, we considered a Markov model and an action set of 26 English letters. The probability transition matrix was distinct for each respondent-item pair and denoted as $\mathbf{P}^{(ij)}= ({p}_{kl}^{(ij)})_{1\leq k,l \leq M}$ for the $i$th respondent and the $j$th item. Given the probability transition matrix, we generated an action sequence starting from ``A'', where the subsequent actions were sampled according to $\mathbf{P}^{(ij)}$ until the final state ``Z'' appears. Excluding the column for ``A'' and the row for ``Z'', the upper right $(M-1) \times (M-1)$ submatrix of $\mathbf{P}^{(ij)}$ was computed according to 
\begin{equation}
{p}_{kl}^{(ij)} = \frac{\exp(\theta_i u_{kl}^{(j)})} { \sum_{r=1}^{M-1}\exp(\theta_i u_{kr}^{(j)})},
\end{equation}
where $(u_{kl}^{(j)})_{1\leq k,l \leq M-1}$ were generated independently from $\mbox{Uniform}(-10,10)$ for each item. 

Two experiments were devised to evaluate the effect of sample size ($N$) and test length ($J$). Experiment I considers four different sample sizes: $N = 200, 500, 1000$ and $2000$. The number of items $J$ was fixed to three with difficulty parameters $b_1 = 0, b_2 = 1, b_3 = -1$ in \eqref{eq:rasch}. Each condition of $N$ was replicated $100$ times. 
Experiment II considers different test lengths. We considered a maximum of twenty items and generated the difficulty parameter $b_j$ from $\mbox{Uniform}(-1,1)$ for each item. Starting from two items, we added one more observed item for estimation in each step until all twenty items were included. The sample size $N$ was fixed at $2000$.

MDS features were extracted from response processes with latent dimension $K$ chosen by five-fold cross-validation from candidate values $10, 20, \ldots, 50$. To guarantee perfect predictability of $Y_j$ by $\mathbf X_j$, the final response to each item was added as an additional dimension to the process features. Following the illustration in Section \ref{sec:procedure}, we first estimated the item parameters $b_1, b_2, b_3$ by marginal maximum likelihood estimation. And then, we used response outcomes to calculate the baseline EAP estimator, $\hat \theta_{\mathbf Y}$, as well as the single-item response-based EAP estimators, $\hat \theta_{Y_1}, \ldots, \hat \theta_{Y_J}$. Note that by using the EAP, we minimize the posterior MSE. 
Ridge regression \citep{tikhonov1977solutions} was used for the first conditional expectation, $E[\hat\theta_{Y_j} | \mathbf X_{-j}]$, and the shrinkage parameter was tuned to minimize the deviance in five-fold cross-validation. For the second conditional expectation, $E[\hat\theta_{\mathbf Y} | T_{\mathbf X_{-j}}, Y_j]$, we regressed $\hat\theta_{\mathbf Y}$ on $(1, T_{\mathbf X_{-j}}, Y_j, T_{\mathbf X_{-j}} Y_j)$ by ordinary least squares.

\subsection{Results}
The estimators $\hat\theta_{\mathbf Y}$ and $\hat \theta_{\mathbf X}$ were evaluated by two criteria, MSE and Kendall's rank correlation \citep[$\tau$;][]{kendall1938new}. The MSE of an estimator $\hat \theta$ was calculated by
\begin{equation}\label{eq:mse}
\text{MSE}(\hat \theta) = \frac{1}{N} \sum_{i=1}^N (\hat\theta_i - \theta_i)^2,
\end{equation}
where $\hat \theta_i$ is the estimate for the $i$-th respondent, and $\theta_i$ is the true latent trait of respondent $i$. The Kendall's $\tau$ between estimated and true $\theta$ can also be calculated for both estimators. In contrast to MSE, Kendall's $\tau$ considers to what extent the estimated ranking aligns with the true ranking of latent trait, which is the interest of norm-referenced tests.

\begin{figure}[h]
	\centering
	\includegraphics[width = \textwidth]{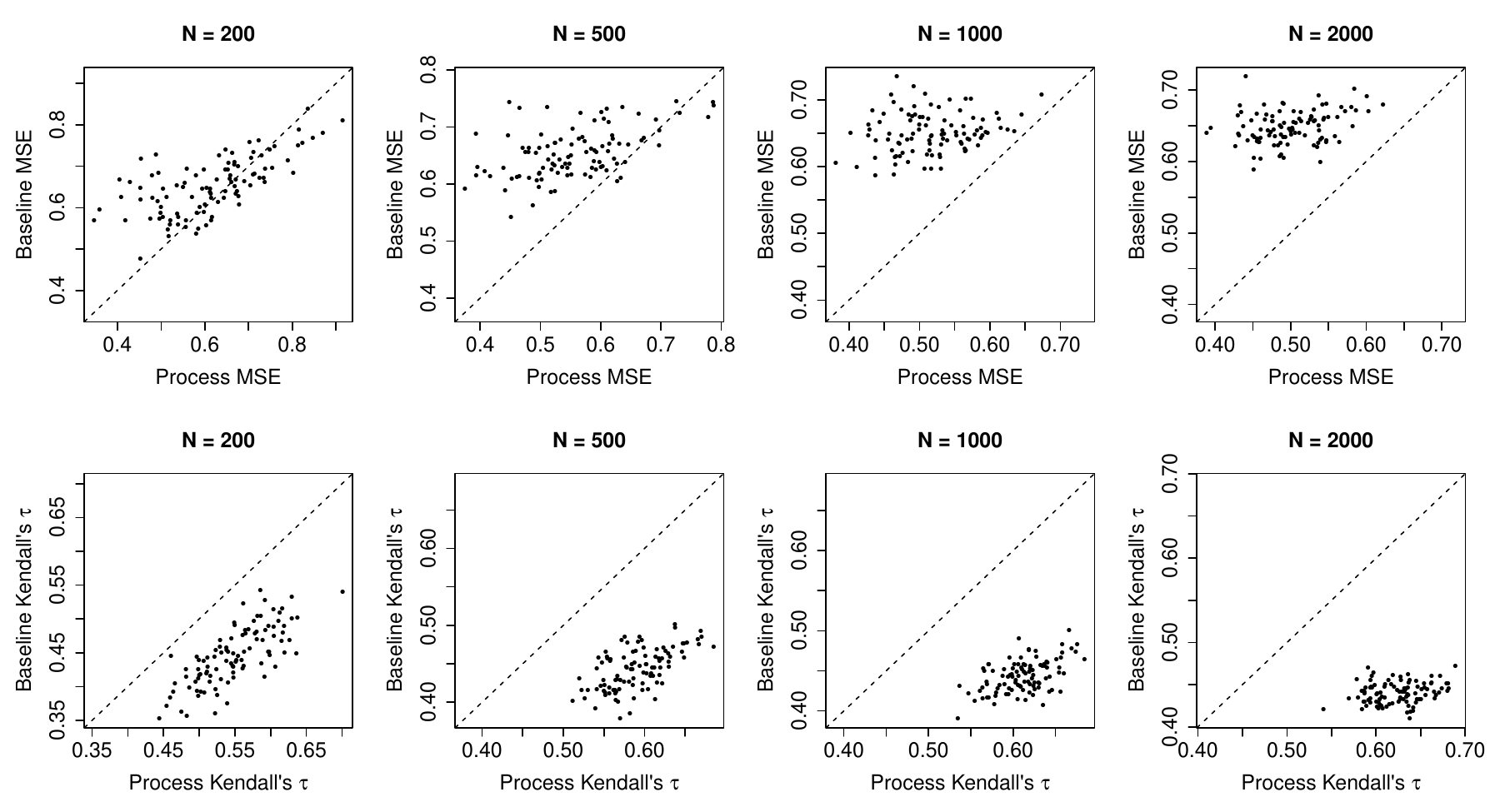}
	\caption{Response- (baseline) and process-based estimators' agreement with true $\theta$ across different sample sizes.}
	\label{fig:sim_partial_sample_size}
\end{figure}
As shown in Figure \ref{fig:sim_partial_sample_size}, the process-based estimator ($x-$axis) outperformed the outcome-based estimator ($y-$axis) in terms of both MSE and Kendall's $\tau$. In each subplot, the 100 points correspond to results from $100$ replications. For smaller sample sizes of $N=200$ or $500$, the MSE of the process-based estimator was higher than that of the response-based estimator in some replications. As sample size increased, the proposed procedure consistently achieved smaller MSE. 
The Kendall's $\tau$ of the process-based estimator was consistently higher than that produced from the baseline estimator across sample sizes and replications, and the improvement became more substantial when $N$ increased.  As shown in the subplot for $N=2000$, $\tau$ could increase from around $0.45$ to over $0.60$ after applying the procedure to incorporate process information.

\begin{figure}[h]
	\centering
	\includegraphics[width = \textwidth]{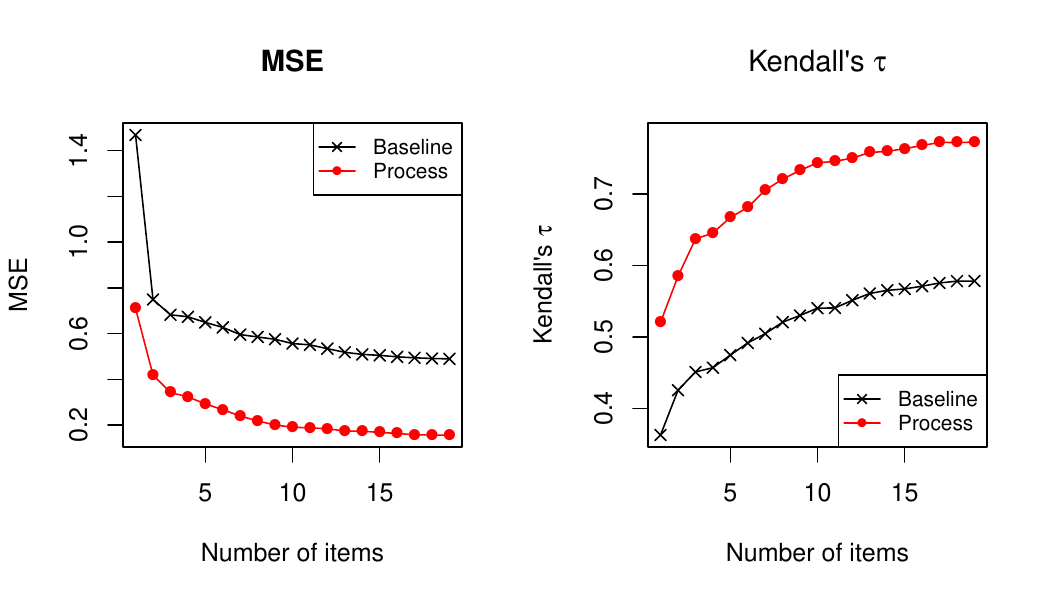}
	\caption{Simulation results of experiment II.}
	\label{fig:step_partial_item_size}
\end{figure}
Figure \ref{fig:step_partial_item_size} displays the results of experiment II where test length was considered. In the left panel, the MSE of the process-based estimator (red line) remained below that of the outcome-based estimator (black line) as the number of items increased from $2$ to $20$. The proposed procedure reduced the MSE by over a half. The improvement in Kendall's $\tau$ was also consistent across different test length as shown in the right panel. For instance, when $J = 7$, Kendall's $\tau$ rose from $0.5$ to $0.7$ after the refinement. 

\section{Empirical Example: PIAAC PSTRE}\label{sec:illustration}

The proposed approach for score refinement was further applied to the data collected from the problem-solving in technology-rich environments (PSTRE) assessment from the 2012 Programme for International Assessment of Adult Competency (PIAAC) survey. The empirical analyses were guided by two overarching objectives. First, the performance of the response- and process-based latent trait estimators were compared, similar to the simulation studies. Second, because response- and process-based estimators are expected to produce different latent ability estimates for the same examinee, we further examined the problem-solving patterns associated with large discrepancies in response- and process-based $\hat\theta$s. In the following subsections, a description of the PIAAC PSTRE data is first provided, followed by the methods and findings from the empirical analyses.

\subsection{The PIAAC PSTRE Data}
Carried out by the Organization for Economic Co-operation and Development (OECD), the PIAAC \citep[e.g.,][]{schleicher2008piaac} is an international survey of the cognitive and workplace skills of working-age individuals around the world. The first cycle of the PIAAC survey in 2012 assessed three cognitive skills, namely literacy, numeracy, and PSTRE, on participants from 24 countries and regions with age between 16 and 65 years. In addition to the three cognitive assessments, the participants were further surveyed on their demographic background and other information related to their occupation and education.

The current study focuses on the PIAAC 2012 PSTRE assessment, where individuals were administered a series of computer-based interactive items. PSTRE ability refers to the ability to use digital technology, communication tools, and internet to obtain and evaluate information, communicate with others, and perform practical tasks \citep{organisation2012literacy}. Successful completion of the PSTRE tasks requires both problem-solving skills and familiarity with digital environments. The test environment of each item resembled commonly seen informational and communicative technology (ICT) platforms, such as e-mail client, web browser, and spreadsheet. Test takers were asked to complete specific tasks in these interactive environments. Individuals' entire log of interactions with each item were recorded as log data. In addition, based on the extent of task completion, polytomous final scores were derived for each item. 

\begin{figure}[h]
    \centering
    \includegraphics[width = \textwidth]{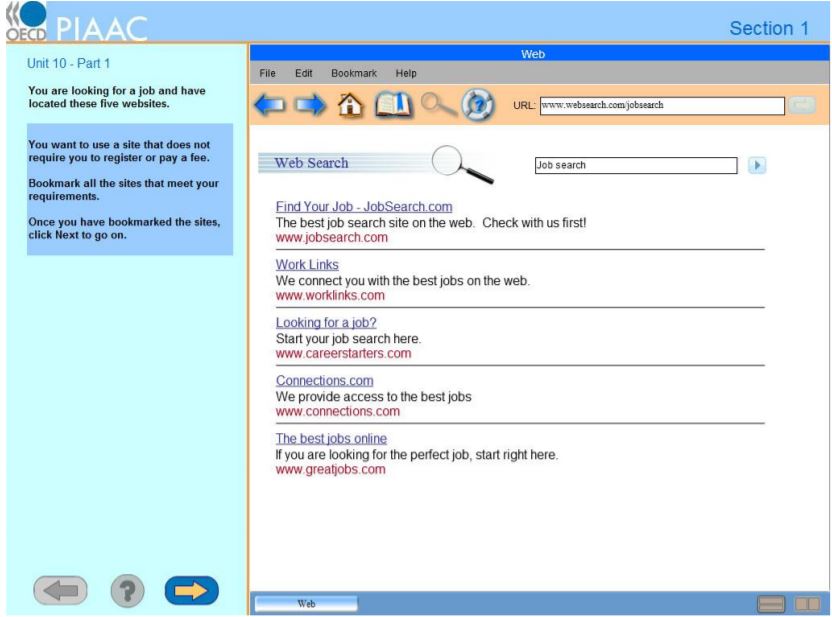}
    \caption{Home page of the PSTRE sample item.} Reprinted from OECD Sample Questions and Questionnaire. 
    \label{fig:sample_item1}
\end{figure}

\begin{figure}[h]
    \centering
    \includegraphics[width = \textwidth]{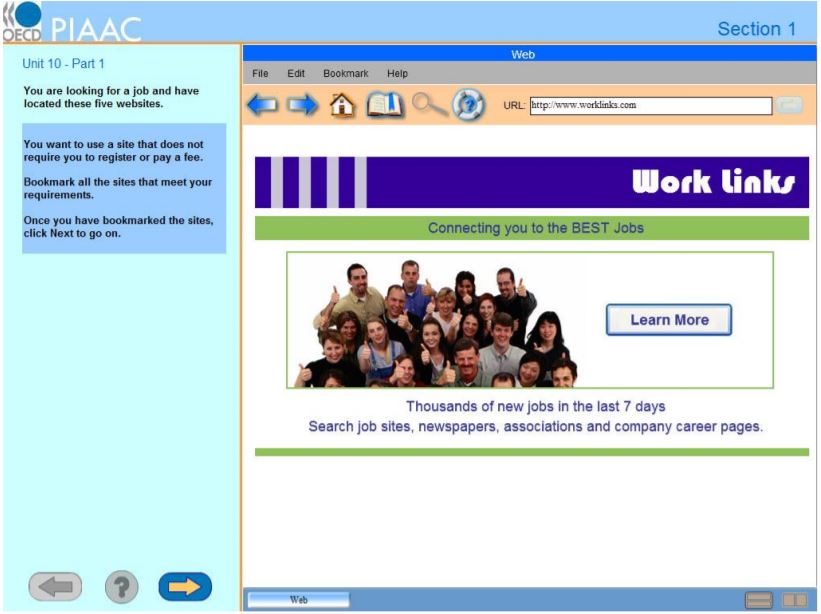}
    \caption{Web page returned from clicking the second link (i.e., ``Work links'') on the home page.}
    \label{fig:sample_item2}
\end{figure}

\begin{figure}[h]
    \centering
    \includegraphics[width = \textwidth]{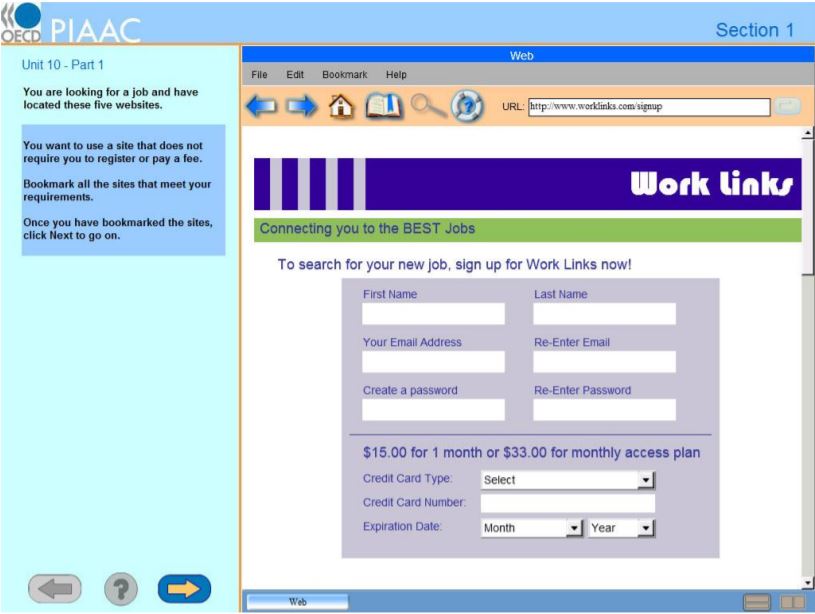}
    \caption{Web page returned from clicking ``Learn More'' on the ``Work links'' website.}
    \label{fig:sample_item3}
\end{figure}

A sample item that resembles PSTRE tasks is shown in Figure \ref{fig:sample_item1}. Respondents can read the task instructions on the left side and work on the task in the simulated interactive environment on the right. This item requires respondents to identify, from the five web pages presented on the screen, all pages that do not require registration or fees and bookmark them. By clicking on each link, they will be redirected to the corresponding website, where they can learn more about the website. For example, clicking ``Work Links'' directs them to Figure \ref{fig:sample_item2}, and further clicking on ``Learn More'' directs them to the page on Figure \ref{fig:sample_item3}. Once having finished working on the task, a test taker can click on the right arrow (``Next'') on the bottom-left. A pop-up window will ask them to confirm their decision by clicking ``OK'' or to return to the question by clicking ``Cancel''. A respondent who clicked on the aforementioned two links, bookmarked the page using the toolbar icon, and moved on to the next question will have the recorded action sequence of ``Start, Click\_W2, Click\_Learn\_More, Toolbar\_Bookmark, Next, Next\_OK''.  

The computer-based version of the 2012 PIAAC survey randomly assigned each respondent with two blocks of cognitive items, where each block consisted of a fixed set of items that assessed either literacy, numeracy, or PSTRE. The current study uses the PSTRE response and process data of individuals from five countries and regions, including the United Kingdom (England and Northern Ireland), Ireland, Japan, the Netherlands, and the United States of America, and who were assigned to PSTRE for both blocks. The five countries and regions were relatively similar in performance distribution on the PIAAC PSTRE assessment. Each PSTRE block consisted of $7$ items, and thus the two blocks total to $14$ items. Note that a recorded action sequence of ``Start, Next, Next\_OK'' indicates that the test taker did not perform any question on the item and moved on to the next question. This type of behavior can be regarded as omission and is distinguished from either credited or uncredited responses. The current study excluded individuals who omitted any of the $14$ items, resulting in a total of $2304$ test takers who responded to all $14$ PSTRE items. For each item, the action sequence of each test taker were recorded, and a polytomous final score calculated based on predefined scoring rubrics was available. These final scores (together with other demographic covariates) were used to estimate individuals' proficiency on PSTRE in the PIAAC survey. Table \ref{tab: PSTRE} presents descriptive information of the $14$ PSTRE items, including the task names and the descriptive statistics of the final scores and action sequences.

\begin{table}
\caption{Descriptives information of the $14$ PIAAC PSTRE items.}
\label{tab: PSTRE}
\begin{tabular}{llcccccc}
  \hline
  &  & \multicolumn{2}{c}{Final Score} & & \multicolumn{3}{c}{Sequence Length}\\ \cline{3-4} \cline{6-8}
  Item ID & Task name & Score levels & Median & Action types & Min & Max & Median \\ 
  \hline
   U01a &  Party Invitations    &  4 &   3 &   40 &   4 &   90 &   17 \\ 
   U01b &  Party Invitations    &  2 &   1 &   47 &   4 &  132 &   29 \\ 
   U02  &  Meeting Room         &  4 &   1 &   95 &   4 &  153 &   35 \\ 
   U03a &  CD Tally             &  2 &   1 &   67 &   4 &   51 &    9 \\ 
   U04a &  Class Attendance     &  4 &   0 &  615 &   4 &  304 &   49 \\ 
   U06a &  Sprained Ankle       &  2 &   0 &   30 &   4 &   57 &   10 \\ 
   U06b &  Sprained Ankle       &  2 &   1 &   26 &   4 &   51 &   18 \\ 
   U07  &  Book Order           &  2 &   1 &   40 &   4 &   79 &   24 \\ 
   U11b &  Locate Email         &  4 &   2 &  122 &   4 &  256 &   22 \\ 
   U16  &  Reply All            &  2 &   1 &  359 &   4 &  267 &   34 \\ 
   U19a &  Club Membership      &  2 &   1 &   75 &   4 &  356 &   19 \\ 
   U19b &  Club Membership      &  3 &   2 &  244 &   4 &  396 &   18 \\ 
   U21  &  Tickets              &  2 &   1 &  124 &   4 &   77 &   22 \\ 
   U23  &  Lamp Return          &  4 &   3 &  133 &   4 &  139 &   25 \\ 
   \hline
\end{tabular}
   \textit{Note.} Descriptive statistics calculated based on the $2304$ participants without omission; Score levels: number of ordinal response categories; Action types: the number of possible actions in the log data; Sequence length: the number of actions performed by a subject.
\end{table}

\subsection{Overall Performance in Latent Proficiency Estimation}

\subsubsection{Evaluation Criteria}

With empirical data, respondents' true $\theta$s were unknown. The two proficiency estimators were instead compared on their agreement with performance on a separate set of items designed to measure the same trait. Specifically, the $14$ PSTRE items were split into two sets of $7$ items. One set of $7$ items, denoted the scoring set ($\mathcal B_{s}$), was used to obtain the response- 
and the process-based estimators ($\hat\theta_{\mathbf Y}^{(s)}$ and $\hat\theta_{\mathbf X}^{(s)}$). A separate latent trait estimate, $\hat\theta_{\mathbf Y}^{(r)}$, can be obtained from the final responses to the remaining $7$ items, denoted the reference set ($\mathcal B_{r}$). Any trait estimate obtained from the scoring set does not use reference set response information, and $\theta_Y^{(r)}$ serves as an external criterion for evaluating $\hat\theta_{\mathbf Y}^{(s)}$ and $\hat\theta_{\mathbf X}^{(s)}$. 
Note that the $14$ items can be partitioned into scoring and reference sets in $\binom{14}{7}$ ways. We randomly chose $50$ possible partitions and evaluated the results on each partition.

Similar to the simulation study, the mean-squared deviation (MSE) with respect to $\hat\theta_{\mathbf Y}^{(r)}$,
\begin{equation}\label{MSE}
    MSE(\hat\theta^{(s)}) = \frac{1}{N}\sum_{i=1}^N (\hat\theta^{(s)} - \hat\theta_{\mathbf Y}^{(r)})^2,
\end{equation}
and the Kendall's $\tau$ with $\hat\theta_{\mathbf Y}^{(r)}$ can be computed for each estimator produced from the scoring set.
Note that, unlike the true $\theta$, $\hat\theta_{\mathbf Y}^{(r)}$ is estimated based on final responses to only $7$ items and contains measurement error. The correlation between $\hat\theta^{(s)}$ and $\hat\theta_{\mathbf Y}^{(r)}$ is hence attenuated by the reliability of $\hat\theta_{\mathbf Y}^{(r)}$, and the MSE of $\hat\theta^{(s)}$ with respect to $\hat\theta_{\mathbf Y}^{(r)}$ is expected to deviate from the MSE of $\hat\theta^{(s)}$ with respect to true $\theta$. Rather than interpreting the two evaluation metrics as the recovery of true proficiency, they can instead be regarded as the split-half ($\mathcal B_s$ and $\mathcal B_r$) agreement of latent trait estimates, or, alternatively, as the strength of association between $\hat\theta^{(s)}$ and performance on similar tasks ($\hat\theta_{\mathbf Y}^{(r)}$). Lower MSE and higher Kendall's $\tau$ hence suggest higher reliability. 

On the scoring set, the response- and process-based estimators were obtained following similar procedures as in the simulation studies. To evaluate performance under different test lengths, similar to experiment II, the number of items in $\mathcal B_s$ used for scoring ranged from $2$ to $7$. Specifically, when only two items were used for scoring, it was assumed that examinees' process and response are observed only on the first two items. Subsequent items in the scoring set were added one by one until all $7$ items were used. Because the final responses were polytomous, the graded response model \citep{samejima2016graded} was used to calibrate the item parameters and to obtain the response-based $\theta$ estimates. Additionally, for the second conditional expectation in Procedure \ref{ch4:procedure:loo}, $E(\hat\theta_{\mathbf Y}\mid T_{\mathbf X_{-j}}, Y_j)$, the polytomous final response $Y_j$ was dummy-coded as regressors in the OLS regression.

\subsubsection{Results}

\begin{figure}[h]
    \centering
    \includegraphics[width = \textwidth]{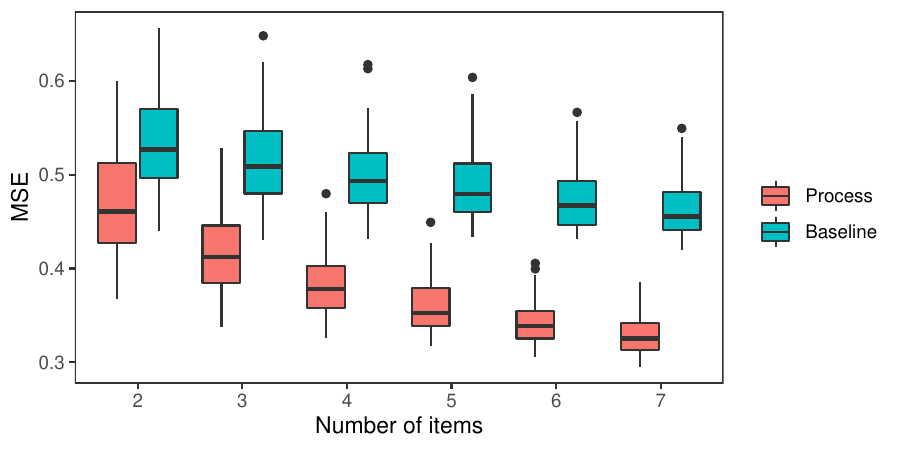}
    \caption{Response- (baseline) and process-based estimators' MSE with respect to reference-set ability estimate on the PIAAC data. $x-$axis is the number of items used for computing $\hat\theta_{\mathbf X}$ and $\hat\theta_{\mathbf Y}$ in the scoring set. Box plots represent the distribution of MSE across the $50$ partitions of $\mathcal B_s$ and $\mathcal B_r$.}
    \label{fig:mse_by_nitems}
\end{figure}

Figure \ref{fig:mse_by_nitems} compares the MSE of the response- and process-based trait estimators. Results are presented for different test lengths, ranging from $2$ to $7$ items, in the scoring set. The green boxes correspond to the response-based (baseline) $\hat\theta^{(s)}$, and the red boxes correspond to the process-based $\hat\theta^{(s)}$. Each box plot in Figure \ref{fig:mse_by_nitems} represents the distribution of the MSE across the $50$ partitions. One can observe that for all test lengths, the process-based latent trait estimator consistently demonstrated smaller MSE, indicating higher agreement with the performance on an external set of similar tasks (i.e., the reference set). 
In particular, with two items, the process-based estimator achieved comparable median MSE with response-based estimator using $5$ items. With four or more items, the process-based $\hat\theta$ consistently achieved similar or lower MSE than the response-based estimator using all $7$ items. 

\begin{figure}[h]
    \centering
    \includegraphics[width = \textwidth]{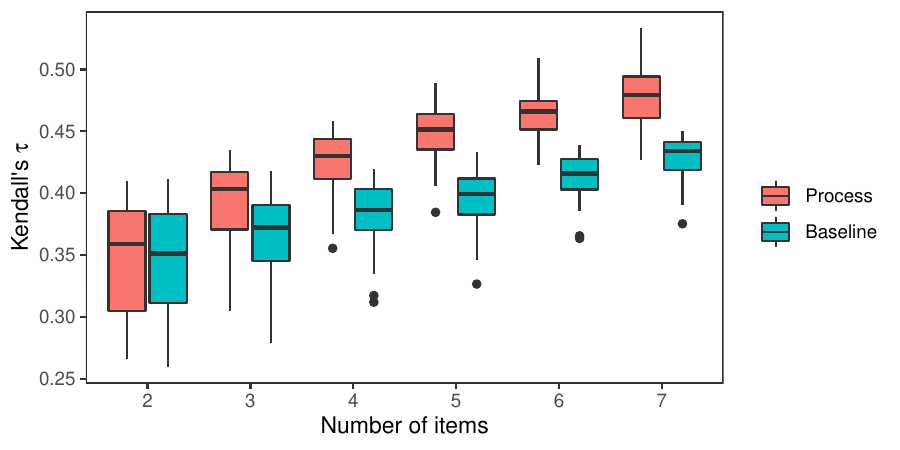}
    \caption{Distribution of response- (baseline) and process-based Kendall's $\tau$ with reference set performance across $50$ partitions on the PIAAC data. }
    \label{fig:kendall_by_nitems}
\end{figure}

The box plots for the Kendall's $\tau$ of the two types of estimators are presented in Figure \ref{fig:kendall_by_nitems}. The correlations with reference set performance were consistently larger using process-based scoring for all test lengths, suggesting that the rankings of latent ability estimates generated based on the problem-solving processes were more similar to the rankings on reference set performance. Again, scores based on processes required less items to achieve a given level of agreement. For instance, with $4$ items, the process-based estimator achieved similar or higher Kendall's $\tau$ when compared to the response-based estimator with all $7$ items. Attenuated by the reliability of the reference set $\theta$ estimate, the absolute Kendall's $\tau$ were mostly below $.5$. When compared to the true $\theta$, however, one would expect the correlation to be higher.

\subsection{Performance by Degree of Process- and Response-based Score Discrepancy}

\subsubsection{Evaluation Methods}

The comparisons above focused on the overall agreement of each estimator with reference set performance. One may also be interested in how the two methods perform for different types of examinees. In particular, it is worth evaluating the relative performance of the two estimators when they disagree on an examinee's latent proficiency ranking. On each of the $50$ partitions, we computed the process- and response-based estimators using all $7$ items on the scoring set. We further regressed the response-based $\hat\theta_{\mathbf Y}^{(s)}$ on the process-based $\hat\theta_{\mathbf X}^{(s)}$ using OLS and computed each individual's Studentized residual for the regression. Individuals were then binned into $10$ groups based on their deciles of the Studentized residuals. The deciles of the Studentized residuals reflect the relative discrepancies in performance ranking based on the two trait estimators: 
For individuals in the first decile, their performance rankings based on process were much lower than that based on responses. Individuals in the $10$th decile, on the other hand, were ranked much higher based on responses than based on process. Individuals closer to the middle ($4$th - $6$th decile) received similar rankings based on process and responses. The MSEs of the two trait estimators with respect to $\hat\theta_{\mathbf Y}^{(r)}$ within each decile were then computed.

\subsubsection{Results}

\begin{figure}[h]
    \centering
    \includegraphics[width = \textwidth]{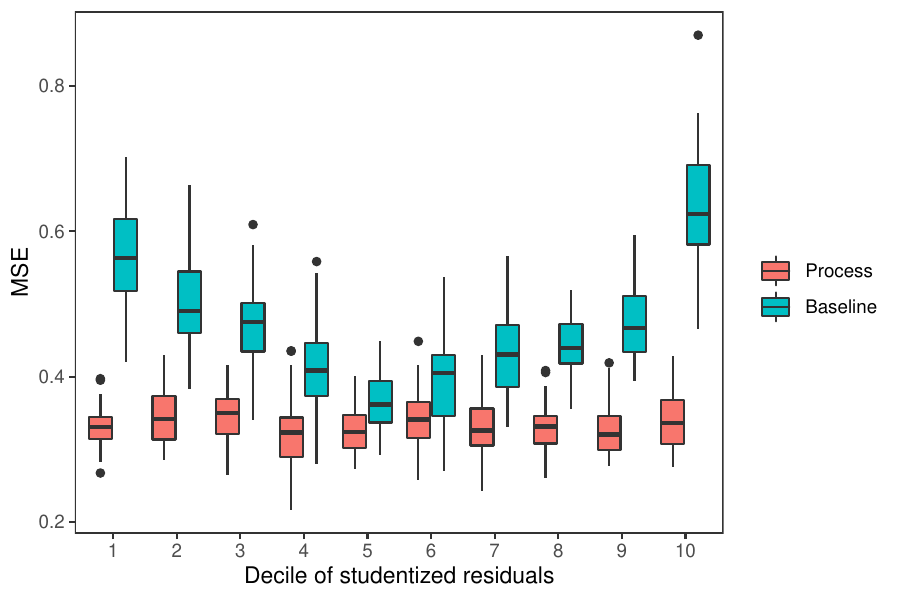}
    \caption{Distribution of MSEs with reference set latent trait estimate in each residual decile. Box plots are the distributions of MSEs within each decile bin across the $50$ partitions of scoring and reference sets.}
    \label{fig:MSE_by_deciles}
\end{figure}

The box plots of the MSEs with respect to reference set performance $\hat\theta_{\mathbf Y}^{(r)}$ across the $50$ partitions, separated by residual deciles, are shown in Figure \ref{fig:MSE_by_deciles}. When the two scores agree on individuals' rankings, the MSEs of $\hat\theta_{\mathbf Y}^{(s)}$ and $\hat\theta_{\mathbf X}^{(s)}$ were similar. However, as we move towards the two ends where the two estimators started to disagree, the MSEs of process-based estimator were remarkably lower than that of response-based estimator. Intuitively, the process- and response-based estimators can be thought of as two judges, one judging individuals' performance based on the problem-solving processes, and the other judging solely based on the final outcome. When the two judges disagree, the process-based estimator consistently better predicted individual performance on similar tasks.

\subsection{Empirical Interpretations of Process- and Response-based Score Discrepancy}

\subsubsection{Methods}

The results above suggested that the proposed process-based latent trait estimate procedures led to an increase in consistency with proficiency estimate on an external set of items, and that the improvement appeared most significant for individuals whose process-based and response-based latent trait estimates disagreed most. One question worth asking is how the proposed approach scores individuals differently compared to the response-based counterpart. We explored this question by looking at the sequences of individuals whose process- and response-based latent trait estimates disagreed the most, i.e., those with the highest or lowest Studentized residuals for $\hat\theta_{\mathbf Y}\sim \hat\theta_{\mathbf X}$. 

This time, with the purpose of interpretation rather than performance evaluation, all $14$ items were used to obtain the two trait estimates. For the individuals in the bottom and top $10$ of the Studentized residuals, we visually examined their action sequences on the $14$ items. 

\subsubsection{Results}

\begin{figure}[h]
    \centering
    \includegraphics[width = 12cm]{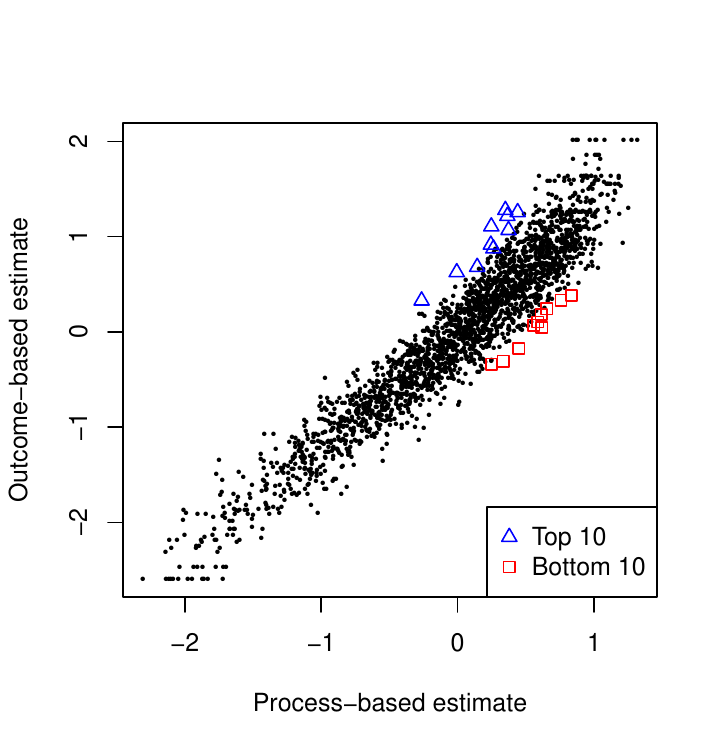}
    \caption{Scatterplot of process- and response-based $\theta$ estimates with $14$ items.}
    \label{fig:scatter_resid}
\end{figure}

Figure \ref{fig:scatter_resid} shows the scatter plot of each respondent's $\hat\theta_{\mathbf X}$ ($x$-axis) and $\hat\theta_{\mathbf Y}$ ($y$-axis). Blue triangles correspond to the $10$ individuals with the highest Studentized residuals, when regressing $\hat\theta_{\mathbf Y}$ on $\hat\theta_{\mathbf X}$. 
These individuals received lower ranking based on processes than based on final responses. For most of them, certain questions were successfully completed but with less efficient strategies: For instance, to look for the requested information from a long spreadsheet, some of these examinees visually inspected every single entry, although a much more efficient strategy is to use ``Search'' or ``Sort'' (e.g., items U03a, U19a). To reply to an email sent to a group, some of them hand-typed the long list of email recipients, when they could simply press ``Reply to all'' (e.g., item U16). Aside from inefficient strategy usage, a few examinees also performed a large number of redundant steps, that is, actions that were not required for successful task completion. 

The red rectangles in Figure \ref{fig:scatter_resid}, on the other hand, represent the $10$ examinees with the lowest Studentized residuals. These examinees received higher ranking based on their problem-solving processes than based on final scores. Several common patterns were observed from their problem-solving processes: The first was partial completion, where the examinee performed some of the key steps on a question, but, before reaching a credited response, proceeded to the next question by clicking ``Next, Next\_OK''. An example is item U16, which required sending an email to a list of recipients containing some key information. Several of the $10$ examinees created the email and filled in the correct content and recipients, but they proceeded to the next question without clicking ``Send''. Another common pattern was careless mistakes, where the examinee demonstrated the required skills for completing the task but slipped on an item due to carelessness. For example, on item U11b, which required sorting emails in the ``Saved'' folder, four of the ten examinees sorted the emails in the ``Inbox'' folder (i.e., default, wrong folder). Intuitively, occasional carelessness and misinterpretation of question requirements, which lead to incorrect responses despite having the requisite skills, may be regarded as one of many sources of random measurement error. With additional information from the problem-solving processes incorporated, the proposed procedure for process-based scoring appeared less impacted by such sources of measurement error.

\section{Discussions}

\subsection{Findings and Implications}
Problem-solving processes contain rich information on individual characteristics, including the measured construct. The current study introduces a method to refine outcome-based latent trait estimates using the additional information from the problem-solving processes. A Rao-Blackwellization approach was proposed for the score refinement. Aside from choosing an appropriate IRT model for the final responses, the proposed approach is relatively data-driven and does not involve prior specification of a measurement model for the problem-solving processes, requiring less subjective inputs compared to expert-defined rubrics for process-based scoring.  

The main theorem states that, under some regularity conditions, the proposed approach can lead to MSE reduction in latent trait estimation. Results from simulation studies corroborate the theorem. An empirical study using the PIAAC PSTRE data further showed that the process-based latent trait estimate tended to have higher agreement with performance on similar tasks, thus higher reliability, compared to the response-based trait estimate. In addition, in order to achieve a particular level of reliability (i.e., MSE or $\tau$ with the external set of items), far fewer items would be required if the additional information from the problem-solving processes is exploited for scoring.

While the improvement in reliability by incorporating process information was consistent across test lengths, one particular merit of incorporating the process information into scoring lies in its use for short tests: With few items, the final responses alone can rarely generate reliable trait estimates. Process information can thus be particularly useful for low-stakes computer-based assessment scenarios, when the administration of long tests is unrealistic or burdensome. With the additional information from problem-solving processes, the tests can be significantly shortened without compromising measurement reliability. An example is interim formative assessments during the learning process, where, after every one or few classes, instructors monitor students' mastery of the recently taught contents. Administration of a long test after each several classes can burden the students and interrupt the learning process. In such cases, a relatively reliable latent ability estimate can be obtained if the problem-solving processes to a few constructed response items are available. Although computerized adaptive testing \citep[CAT; e.g.,][]{wainer2000computerized} also reduces test length through the adaptive selection of test items tailored to individuals real-time performance, the construction of a CAT usually requires a large pre-calibrated item pool with hundreds of items, which may be hard to achieve for small-scale, low-stakes assessments. The production of a process-based scoring rule, on the other hand, only requires sufficient items for reliable measurement of latent proficiency and a sample size sufficient for item parameter calibration, process feature extraction, and training the conditional expectation models. 

\subsection{Practical Considerations}

The proposed approach could consistently improve test reliability. However, there are a few caveats for its implementations, especially for exams with higher stakes. First, in the empirical study, the performance of the process- and response-based latent trait estimators were evaluated using up to $7$ items for scoring. The choice of up to $7$ items was due to the limited number of total items available ($14$) and the need to set aside a large enough reference set of items used for evaluations. For an operational test, however, $7$ items' final responses are far from sufficient for reliable measurement, and the measurement error in the final response-based latent trait estimates can propagate to the process-based scores through the conditional expectation. Test developers are advised to have a sufficiently large scoring set, so that relatively reliable response-based latent trait estimates can be obtained. Second, the proposed process-based scoring approach aimed at improving the measurement precision, or reliability, of the assessment through MSE reduction. The validity of the scoring rule, however, is a separate critical issue to be addressed. Looking at the empirical interpretations of the process-based scores, it appeared that individuals were scored higher based on processes when they gave up on the track to a correct response, demonstrated partially correct responses, or slipped on the final response due to careless mistakes. In these cases, increasing the individuals' latent trait estimates may be reasonable, because each of these patterns demonstrated partial or full mastery of the required skills for completing the tasks. Meanwhile, individuals who reached correct responses but with less efficient problem-solving strategy received lower process-based proficiency estimate. Assigning different trait estimates based on the choice of test-taking strategies may be more controversial: In the empirical study, the use of more efficient problem-solving strategy was found positively correlated with the final score on other tasks, providing validity evidence on the use of such information for ability assessment. From this perspective, problem-solving strategy information may be incorporated into ability estimation similar to other types of collateral information, such as response times \citep[e.g.,][]{van2007hierarchical,bolsinova2018improving} and other covariates in latent regression \citep[e.g.,][]{von200632}. On the other hand, test takers may be unaware that they are scored based on other information aside from task completion, raising concerns on the face validity and broader implications of the scoring criterion. At the same time, it raises test design questions such as whether examinees should be informed that their score can be affected by their problem-solving process. Evaluation of measurement validity would involve not only a search of empirical evidence that support the use and interpretation of the scores but also the appraisal of social consequences of score uses \citep{messick1989meaning}. We leave the question of how to best assist experts with the validation of data-driven latent proficiency estimators to future research.

To establish the theoretical results on improved measurement efficiency, several assumptions were made in the current framework. One should note that, if some of the assumptions are violated, the efficiency results may be discounted, and several sources of bias may be introduced to the estimator. In particular, assumption A1 is the most important among all. It provides a crucial guide to extract the relevant part of the process data for the assessment. This also assumes that the final-response-based IRT model yields a valid estimator with noise, written as $\hat \theta_{Y_j,i} = m_j(\theta_i) + \varepsilon_i$. Assumption A2 ensures that $\varepsilon_i$ is not predictable by the process data of other items and furthermore $\varepsilon_i$ has zero mean across the population. Thus, if A2 is not valid, we may expect certain amount of bias introduced to the process estimator. A3 is a technical framework for us to discuss efficiency. We choose natural exponential family because it is the first order approximation of a large class of parametric families. We do expect that the proposed estimator improves upon the classic IRT estimator even beyond the these three assumptions. In practice, we have not developed practical diagnostic procedures to verify these assumptions by the data. However, if there are multiple items, practitioner could also perform an analysis as was performed in the real data analysis to compare the process-data-based score and the IRT-model-based score, which is the ultimate goal of this analysis.

\subsection{Future Extentions}

The methods for data-driven score refinement based on problem-solving processes can be extended in several ways. To start, several assumptions were made in the current theoretical framework, including the conditional independence between an item's final response and the process on other items, as well as the exponential family distribution of the process features given the measured trait. Statistical methods for empirically testing these assumptions should be developed for the practical implementation of the proposed method. Second, while the current study provides one approach to increasing measurement reliability with process information, other methods, such as latent regression with process as covariate and confirmatory models with both response and process indicators, may be developed. Unlike the ordinal final outcomes which are designed to assess the measured trait, the problem-solving process data is high dimensional and can contain substantial construct-irrelevant variance. For these parametric models, effective methods for variable selection will be needed to parse out the signal ($\theta$-related information) from the ``noise''. Another potential extension of the process-based scoring method is to diagnostic assessments \citep[e.g.,][]{templin2010diagnostic}, where, instead of measuring individuals on the continuous proficiency continuum, the goal is to classify individuals into latent classes based on their mastery status of discrete skills.

\vspace{\fill}\pagebreak

\bibliography{process_Ref}

\newpage
\section*{Appendix: Proofs of Theorem 1 and Theorem 2}
	To prove Theorem \ref{ch4:thm:sufficient_stat}, we establish the following lemma.
\begin{lemma}\label{ch4:lemma:cov_monotone}
	Let $X$ be a nonconstant random variable, and  $f(\cdot)$ and $g(\cdot)$ be strictly increasing functions. Suppose that $f(X)$ and $g(X)$ have finite second moments. Then $\cov \left( f(X), g(X) \right) >0$ .
\end{lemma}

\begin{proof}[Proof of lemma \ref{ch4:lemma:cov_monotone}]
	Let $Y$ be an independent and identically distributed (i.i.d.) copy of $X$. It is easy to verify the following identity	\begin{equation}\label{ch4:ineq: monotone}
	\cov \left( f(X), g(X) \right) = \frac{1}{2} E \left[ \left(f(X) -f(Y) \right) \left(g(X) - g(Y) \right)  \right].
	\end{equation}
	Clearly, for any $x$ and $y$, $ (f(x) -f(y) ) (g(x) - g(y))\ge 0$, and $``=''$ holds if and only if $x=y$. Since $P(X\not=Y)>0$, the right-hand side of equation \eqref{ch4:ineq: monotone} must be positive. 		
\end{proof}

\begin{proof}[Proof of Theorem \ref{ch4:thm:sufficient_stat}]
	By Assumption A2 (local independence), 
	\begin{equation*}
	T_{\mathbf X_{-j}} = E \left[\hat \theta_{Y_j} | \mathbf X_{-j}\right]  
	= E \left[ E \left[\hat\theta_{Y_j} | \mathbf X_{-j} ,\theta \right] | \mathbf X_{-j} \right]
	= E \left[ E \left[\hat\theta_{Y_j} | \theta \right] | \mathbf X_{-j} \right]  =  E \left[ m_j (\theta) | \mathbf X_{-j} \right].
	\end{equation*}
	Due to Assumption A3 (exponential family), the posterior distribution of $\theta$ given $\mathbf X_{-j}$ depends on $\mathbf X_{-j}$ only through the sufficient statistic $T_j(\mathbf X_{-j})$. In fact, 
	\begin{equation*}
	T_{\mathbf X_{-j}} = E \left[ m_j (\theta) | \mathbf X_{-j} \right] = G_j(T_j(\mathbf X_{-j})),
	\end{equation*}
	where $G_j(t) =E \left[ m_j (\theta) |T_j(\mathbf X_{-j})=t\right]$. Furthermore, by making use of the exponential family form in Assumption A3 and the simple exchange of order of differentiation and integration, we can show that  	
	\begin{equation*}
	G_j'(t) =  \cov\left[m_j(\theta),\eta_j(\theta) | T_j(\mathbf X_{-j})=t \right].
	\end{equation*}
	Since both $m_j$ and $\eta_j$ are strictly monotone,  Lemma \ref{ch4:lemma:cov_monotone} implies that $G_j'(t)$ is strictly positive or negative for all $t$ and, therefore, $G_j$ is strictly monotone. In other words, there is a one-to-one mapping between $T_{\mathbf X_j}$ and $T_j(\mathbf X_{-j})$.
\end{proof}

\begin{proof}[Proof of Theorem \ref{ch4:thm:var_reduction}]
	From Theorem \ref{ch4:thm:sufficient_stat}, we know that $T_{\mathbf X_{-j}}$ is a sufficient statistic of $\mathbf X_{-j}$ for each $j$. Since $\hat\theta_{\mathbf Y}$ is a function of $\mathbf Y$ and $\sigma(\mathbf Y_{-j}) \subseteq \sigma(\mathbf X_{-j})$, the conditional distribution $\hat\theta_{\mathbf Y} | T_{\mathbf X_{-j}}, Y_j$ is free of $\theta$. Therefore, we have 
	$E[\hat \theta_{\mathbf Y} | T_{\mathbf X_{-j}}, Y_j, \theta] = E[\hat\theta_{\mathbf Y} | T_{\mathbf X_{-j}}, Y_j ] = \hat \theta_{\mathbf X_{-j}}.$
	It follows from the well-known Rao-Blackwell Theorem \citep{casella2002statistical} that $\hat \theta_{\mathbf X_{-j}}$ reduces the conditional variance and 
	$$ E[(\hat \theta_{\mathbf X_{-j}} - \theta)^2 | \theta] \leq E[(\hat \theta_{\mathbf Y} - \theta)^2 | \theta]$$ holds for every $j$ and $\theta$.
	By Cauchy-Schwarz inequality, we get 
	$$E[(\hat \theta_{\mathbf X} - \theta)^2 | \theta] \leq  E [\frac{1}{J} \sum_{j=1}^J (\hat \theta_{\mathbf X_{-j}} - \theta)^2 | \theta] \leq E[(\hat \theta_{\mathbf Y} - \theta)^2 | \theta].$$
\end{proof}

\end{document}